# Unidirectional focusing of light using structured diffractive surfaces

Yuhang Li[1,2], Tianyi Gan[1,2], Jingxi Li[1,2], Mona Jarrahi[1,2] and Aydogan Ozcan[*,1,2,3]

[1]Electrical and Computer Engineering Department, University of California, Los Angeles, CA, 90095, USA.

[2]California NanoSystems Institute (CNSI), University of California, Los Angeles, CA, USA.

[3]Bioengineering Department, University of California, Los Angeles, 90095, USA.

[*]Correspondence: Aydogan Ozcan. Email: ozcan@ucla.edu

## Abstract

Unidirectional optical systems enable selective control of light through asymmetric processing of radiation, effectively transmitting light in one direction while blocking unwanted propagation in the opposite direction. Here, we introduce a reciprocal diffractive unidirectional focusing design based on linear and isotropic diffractive layers that are structured. Using deep learning-based optimization, a cascaded set of diffractive layers are spatially engineered at the wavelength scale to focus light efficiently in the forward direction while blocking it in the opposite direction. The forward energy focusing efficiency and the backward energy suppression capabilities of this unidirectional architecture were demonstrated under various illumination angles and wavelengths, illustrating the versatility of our polarization-insensitive design. Furthermore, we demonstrated that these designs are resilient to adversarial attacks that utilize wavefront engineering from outside. Experimental validation using terahertz radiation confirmed the feasibility of this diffractive unidirectional focusing framework. Diffractive unidirectional designs can operate across different parts of the electromagnetic spectrum by scaling the resulting diffractive features proportional to the wavelength of light and will find applications in security, defense, and optical communication, among others.

## Introduction

The control of asymmetric light propagation, where light preferentially travels in one direction while being suppressed or scattered in the opposite direction, has enabled various advancements in optical systems. Achieving such directional control of light propagation is crucial for applications in wave transmission, imaging, and sensing[1–3]. However, most optical systems based on linear and time-invariant components are inherently bidirectional. Breaking this to enable asymmetric light propagation presents some challenges. Traditional approaches often rely on advanced material properties, such as the magneto-optic effect[4–7] or nonlinear materials[8–10]. These methods, while effective, are typically associated with relatively bulky and costly setups due to their tight fabrication requirements and rely on high-power laser sources to introduce sufficient nonlinearity. Additionally, asymmetric isotropic dielectric gratings[11–13] and metamaterials[14–18] have been explored to create unidirectional material systems. Although these approaches demonstrated success, such systems are often limited in scope due to their complex design and fabrication processes, polarization sensitivity and poor performance under off-axis illumination.

In this work, we report a reciprocal diffractive optical system designed for unidirectional focusing of radiation, where the incoming light is focused in the forward direction while being blocked or scattered in the opposite direction, as illustrated in Fig. 1. This unidirectional propagation is achieved using structurally optimized linear and isotropic diffractive layers with wavelength-scale features that are optimized using deep learning. The unidirectional system's energy-blocking capability in the backward direction was evaluated under multi-angle illumination. Incorporating various illumination angles during the optimization process enhanced the system's light-blocking performance in the backward direction, enabling it to generalize effectively across different angles of illumination. Furthermore, this training strategy also improved the unidirectional system's resilience against adversarial attacks, making it impractical to focus light in the backward direction regardless of any wavefront engineering from outside. Additionally, we extended our polarization-insensitive unidirectional focusing designs to operate across multiple wavelengths, demonstrating the ability to maintain unidirectional light control within a broad spectral range of interest. The proof of concept of this unidirectional focusing framework was experimentally validated using 3D-printed diffractive layers that operate at the terahertz part of the spectrum. We believe that the presented diffractive unidirectional focusing architecture will find various applications in defense/security, optical communication, imaging and sensing.

## Results

**Unidirectional Focusing of Light Using Structured Diffractive Materials**

Our framework for unidirectional focusing of coherent light is illustrated in Fig. 1. Light entering the system through an input aperture, i.e., the input field of view A (FOV A), is focused onto a specific output region/aperture within FOV B (A→B defining the forward focusing operation, Fig. 1a), whereas the light originating from the aperture in FOV B is scattered out (B→A defining the backward operation, Fig, 1b). As shown in Fig. 2a, the unidirectional focusing design consisted of four phase-only diffractive layers, each containing 240 × 240 phase-modulating features, approximately $\lambda/2$ in lateral size, where $\lambda$ is the illumination wavelength. The phase modulation of each diffractive feature ranges from 0 to $2\pi$, achieved by adjusting its local thickness (refer to the Methods section for details). During the design optimization process, the phase values of the diffractive layers were iteratively updated using a stochastic gradient descent-based algorithm and a custom loss function. This loss function was mathematically constructed to maximize the output energy efficiency in the forward path (A→B) while minimizing the energy efficiency in the backward path (B→A). After its training, the system can successfully focus a plane wave passing through an aperture in FOV A into a tightly focused spot at the output (FOV B) within a diameter of $4.3\lambda$, while effectively blocking the light traveling in the reverse direction, B→A. As shown in Figs. 2b and 2c, the forward focusing operation achieved ~98.7% energy efficiency within the target output region (FOV B), whereas the backward illumination resulted in only ~0.4% energy detected in FOV A.

The phase modulation patterns of the optimized diffractive layers are visualized in Fig. 2d. One can observe that the diffractive layers exhibit an asymmetric structure from A to B vs. B to A, and the central regions of the resulting diffractive layers exhibit lens-like structures, which are crucial for the efficient focusing of light in the forward operation. However, the edge regions of each diffractive layer adopt grating-like structures that scatter and diffuse light during the backward operation, enhancing beam suppression in the reverse illumination path.

These numerical analyses underscore the effectiveness of phase-only diffractive designs in achieving efficient unidirectional focusing of light through structural asymmetry. By leveraging customized loss functions and gradient descent-based optimization, this approach enables diffractive systems to selectively control light propagation properties in different directions.

**Robustness of Diffractive Focusing Systems Under Oblique Illumination and Adversarial Attacks**

While our diffractive focusing system reported in Fig. 2 exhibits a very good unidirectional performance under normal illumination, real-world scenarios often involve oblique incidences of incoming radiation. To evaluate the robustness of our design, we analyzed its behavior under oblique illumination. Specifically, we tested the backward operation (B→A) of the system in Fig. 2 using plane waves incident at various illumination angles relative to the optical axis, as shown in Fig. 3a. The phase distributions were adjusted to control the incidence angle, maintaining uniform amplitude within the aperture of FOV B. Initially, the system was trained exclusively with plane waves propagating normal to the input aperture. While this approach effectively blocked the reverse beam propagation (B→A) for the normal incidence of illumination, its performance deteriorated as the incidence angle increased in the testing phase. For illumination angles approaching ~20°, the suppression capability of the diffractive system for B→A significantly degraded, allowing as much as 60% of the energy to leak into FOV A in the backward direction (see Fig. 3b).

To address this performance limitation under oblique illumination, we implemented a multi-angle training strategy that incorporated diverse illumination conditions during the optimization of the cascaded diffractive layers. Forward operation (A→B) training remained limited to normal incidence, but the backward operation (B→A) was trained using various oblique waves randomly selected within a specified illumination range $[\theta_{min}, \theta_{max}]$. This approach allowed the diffractive system to generalize its beam suppression response in the backward direction across a broad range of oblique illumination angles by optimizing the phase modulation patterns of the diffractive layers. As shown in Fig. 3c, this multi-angle-trained system demonstrated substantial performance improvements. During its training, $[\theta_{min}, \theta_{max}]$ was selected as $[-20°, 20°]$ (red box in Fig. 3c), while the blind testing of its performance was conducted over a broader range of illumination angles covering $[-40°, 40°]$. Compared to the original diffractive system trained only for the normal incidence of illumination, the multi-angle-trained diffractive unidirectional system showed significant improvements in its backward beam suppression capability across a wide range of angles, with only a minor reduction in the forward beam focusing efficiency, dropping to 97.2% from ~98.7% of the previous design. Notably, even for the incident angles beyond the training range, the diffractive unidirectional system maintained a good beam suppression capability in the reverse direction. In this improved design, the trained diffractive layers, shown in Fig. 3d, featured more complex patterns, particularly around the edge regions, which contribute to enhanced optical scattering and losses during the backward operation with oblique illumination angles.

Another critical consideration is the potential vulnerability of the unidirectional beam-focusing system to *adversarial* manipulations from the outside. For example, if an adversary from outside has access to the optical signal at FOV A through a "spying" detector, they could optimize the structure of the incident phase at FOV B by manipulating the wavefront in the backward direction (B→A) to gradually increase the energy at the output FOV A. To assess this vulnerability from the outside, we performed phase optimization on the incident wave in the backward path using iterative feedback from the optical signal observed at FOV A (see the Methods section for details). As shown in Fig. 4a, diffractive unidirectional systems trained exclusively with the normal incidence of illumination were highly susceptible to such adversarial attacks, with the energy leakage in the backward direction reaching ~88% through feedback-based learning using a "spying" detector at FOV A. This high energy leakage under an optimized attack underscores the limited robustness of these systems to adversarial phase manipulation from outside. In contrast, the multi-angle-trained diffractive unidirectional system reported in Fig. 3 exhibited much better resilience to such adversarial attacks. By optimizing the phase modulation patterns for a range of incidence angles, the diffractive layers exhibited more complex and resilient structures, making it much harder for an adversary to find an optimal phase configuration to "hack" the system in the reverse path, B→A. Consequently, the resulting energy efficiency in the backward path was significantly lower than it was in the vanilla diffractive unidirectional systems. This increased robustness against adversarial attacks from outside underscores the advantages of training diffractive unidirectional systems with diverse illumination conditions, enhancing not only their generalization capabilities but also their resilience to targeted attacks from outside. Expanding the training process to include an even wider range of illumination angles and other wavefront perturbations can ensure stronger energy suppression in the backward path, even under challenging or adversarial conditions. Such strategies are critical for developing practical and reliable diffractive unidirectional focusing systems for real-world applications.

Finally, we should emphasize that in the adversarial attacks that were considered here, we assumed that there was direct access to the beam power at FOV A in order to iteratively optimize the attacking wavefront in the reverse direction; without such an internal feedback mechanism through, e.g., a "spy" from inside, an adversary from outside would not be able to find any practical method to send radiation in the backward direction unless the design of the diffractive unidirectional material and its layers are a priori known.

**Multi-Wavelength Operation**

We further demonstrate that the diffractive focusing framework can operate at multiple illumination wavelengths, as shown in Fig. 5a. The system, configured similarly to the previously shown monochromatic unidirectional design, was trained using three illumination wavelengths: 0.7, 0.75, and 0.8 mm. After its training, the diffractive unidirectional focusing design achieved forward focusing efficiencies of 92.42%, 93.46% and 91.69% for the three illumination wavelengths, respectively. In the backward operation, the system effectively blocked the input light, with a leakage of 2.79%, 1.36%, and 1.71% at these 3 wavelengths, 0.7, 0.75, and 0.8 mm, respectively. However, this multi-wavelength operation comes with a slight performance trade-off. Compared to the monochromatic unidirectional design, the multi-wavelength system exhibited a small reduction in the peak forward diffraction efficiency. For a given number of trainable diffractive degrees of freedom, this behavior reflects the inherent balance between optimization of the unidirectional focusing performance for a specific wavelength vs. achieving broader spectral functionality. These reported multi-wavelength performance metrics can be further improved using deeper diffractive designs with more degrees of freedom available for optimization.

We also blindly evaluated the performance of the diffractive unidirectional focusing system under illumination across a range of wavelengths. For the system trained with monochromatic illumination at $\lambda = 0.75$ mm, the unidirectional focusing performance degraded significantly as the wavelength deviated from the training wavelength, as shown with the blue and red lines in Fig. 5c. In the forward direction (A→B), the diffraction efficiency dropped sharply outside the optimal spectral range that the diffractive system was trained for. Similarly, in the backward operation (B→A), the energy leakage percentage increased, particularly for wavelengths shorter than 0.75 mm.

In contrast, the diffractive unidirectional focusing design trained with three distinct wavelengths (0.7, 0.75, and 0.8 mm) demonstrated far superior spectral performance. It maintained over >70% forward focusing efficiency across a broader spectral range while suppressing the backward illumination to <5% from 0.690 mm to 0.814 mm illumination. As indicated by the yellow and purple lines in Fig. 5c, the multi-wavelength training significantly improved the system's output efficiency for wavelengths not explicitly used during the training – which indicates the *external generalization* behavior of the diffractive unidirectional processor outside of its training spectral range.

**Experimental Demonstration of a Diffractive Unidirectional Focusing System**

We experimentally validated our diffractive unidirectional focusing design using monochromatic continuous-wave terahertz illumination at λ = 0.75 mm, as shown in Fig. 6. The schematic diagram of the terahertz setup is presented in Fig. 6a, with implementation details reported in the Methods section. For this experimental validation, we designed a diffractive unidirectional focusing system consisting of two diffractive layers, each composed of $120 \times 120$ learnable diffractive features, each with a lateral size of $0.64\lambda$. The axial spacing between the adjacent planes was set to ~26.7λ. The aperture size of FOV A was $42.67\lambda$, while the aperture of FOV B was $6.4\lambda$ (at the output focus plane). Unlike previous designs that assumed phase-only diffractive layers, this model incorporated the complex-valued refractive index of the 3D-printing diffractive material to account for material absorption, ensuring that optical absorption by the layers was mathematically taken into account during the deep learning-based design process.

After the optimization phase, the resulting diffractive layers were fabricated using 3D printing, as shown in Fig. 6c. The experimental performance of the fabricated system was evaluated for both the forward and the backward operations, as depicted in Fig. 6d. The results demonstrated that the 3D-printed diffractive unidirectional focusing system successfully focused light within the target output region in the forward direction (A→B) while effectively scattering light in the backward direction (B→A). These results demonstrate the experimental feasibility of the presented diffractive unidirectional focusing framework.

## Discussion

In this work, we introduced a diffractive unidirectional focusing framework, showcasing the potential of all-optical diffractive processors for asymmetric light manipulation. The system demonstrated efficient forward light focusing while effectively suppressing energy leakage in the backward direction. By incorporating multi-angle illumination during the training, the system's robustness was enhanced, allowing it to maintain unidirectional performance under oblique random illumination. This diverse training approach also improved its generalization capabilities, mitigating energy leakage from oblique incident waves in the backward direction, and further increased resilience against adversarial attacks from outside. Additionally, we explored the unidirectional system's spectral response and broadband performance, showcasing the effectiveness of involving multiple wavelengths during the training phase to adapt the system to work under a broader illumination spectrum. Moreover, the feasibility of this unidirectional focusing framework was validated experimentally using monochromatic terahertz illumination, confirming its practical implementation.

Although the presented designs are based on spatially coherent illumination, this framework can be extended to spatially incoherent or partially coherent input fields by leveraging the same design principles and deep learning-based optimization methods[19–21]. For spatially incoherent or partially coherent illumination, phase-only diffractive layers can be optimized using similar unidirectional focusing-related loss functions, although simulating incoherent or partially coherent field propagation requires additional steps to statistically account for the coherence diameter at the illumination plane, which relatively increases computational demand and training time. Extending these systems to operate effectively in various environments with complex, incoherent or partially coherent light sources will further broaden their applicability, expanding the utility of these designs for further innovations in unidirectional optical systems.

## Methods

**Optical Forward Model of a Unidirectional Focusing System**

In the forward model of our diffractive unidirectional focusing design, the input plane, diffractive layers, and output plane are sequentially arranged along the optical axis, with an axial spacing of $d$ between each plane. For the numerical and experimental models, $d$ was empirically set to $6\ mm$ and $20\ mm$, respectively, corresponding to $8\lambda$ and $26.67\lambda$, where $\lambda = 0.75$ mm is the operating wavelength. The optical forward model of a diffractive processor consists of two sequential processes: (1) free-space propagation of the light wave between consecutive planes and (2) modulation of the light wave by the apertures or the diffractive layers. The free-space propagation was modeled using the angular spectrum approach[22], expressed as:

$$u(x,y,z+d) = \mathcal{F}^{-1}\{\mathcal{F}\{u(x,y,z)\} \cdot H(f_x,f_y;d)\} \quad (1),$$

where $u(x,y,z)$ represents the complex-valued field at a coordinate of $z$ along the optical axis, and $u(x,y,z+d)$ is the resulting field at the coordinate of $z+d$ after propagating over an axial distance of $d$. Here, $f_x$ and $f_y$ denote the spatial frequencies along the $x$ and $y$ directions, respectively, while $\mathcal{F}$ and $\mathcal{F}^{-1}$ denote the 2D Fourier transform and 2D inverse Fourier transform, respectively. The free-space transfer function $H(f_x,f_y;d)$ is defined as:

$$H(f_x,f_y;d) = \begin{cases} \exp\left\{jkd\sqrt{1-\left(\frac{2\pi f_x}{k}\right)^2 - \left(\frac{2\pi f_y}{k}\right)^2}\right\}, & f_x^2 + f_y^2 < \frac{1}{\lambda^2} \\ 0, & f_x^2 + f_y^2 \geq \frac{1}{\lambda^2} \end{cases} \quad (2),$$

where $j = \sqrt{-1}$ and $k = \frac{2\pi}{\lambda}$.

The diffractive layers were modeled as thin optical modulation elements. The $m$-th diffractive feature on $k$-th layer, located at $(x_m, y_m, z_m)$, is described as:

$$t_k(x_m, y_m, z_m) = \exp\{j\phi_k(x_m, y_m, z_m)\} \quad (3),$$

where $\phi_k(x_m, y_m, z_m)$ represents the phase modulation value for the corresponding diffractive feature.

In the experimental demonstration, the height values of the learnable diffractive features ($h_{trainable}$) were computed based on the refractive index of the 3D printing material[23]. Additionally, a constant substrate thickness ($h_{substrate} = 0.5\ mm$) was added for mechanical support of each layer.

**Experimental Demonstration**

The diffractive layers were fabricated using a 3D printer (Objet30 Pro, Stratasys). The apertures were also 3D printed and coated with aluminum foil to define the light-blocking regions, while the uncovered areas served as the transmission zones. A 3D-printed holder was used to assemble the diffractive layers and input objects in alignment with the positions specified in the numerical design.

A terahertz continuous-wave scanning system was used for testing our diffractive unidirectional focusing design. As depicted in Fig. 6b, the terahertz source comprised a WR2.2 modular amplifier/multiplier chain (AMC) with a compatible diagonal horn antenna (Virginia Diodes Inc.). The AMC received a 10-dBm radiofrequency (RF) input signal at 11.1111 GHz ($f_{RF1}$), which was multiplied 36 times to generate output radiation at 400 GHz, corresponding to a wavelength of λ = 0.75 mm. The AMC output was modulated with a 1-kHz square wave for lock-in detection. The assembled diffractive unidirectional focusing system was positioned approximately 600 mm from the horn antenna's exit aperture, ensuring a near-uniform plane wave illumination across its input field of view (FOV A), with dimensions of $36 \times 36\ mm^2$ (i.e., 48λ × 48λ). The intensity distribution within the output FOV (B) of the imager was scanned with a step size of 0.8 mm using a single-pixel detector system. The detector comprised a mixer/AMC (Virginia Diodes Inc.) mounted on an $xy$-positioning stage, assembled from two linear motorized stages (Thorlabs NRT100). The detector also received a 10-dBm sinusoidal local oscillator signal at 11.083 GHz ($f_{RF2}$) for mixing to down-convert the output signal to 1 GHz. The down-converted 1-GHz signal was amplified by a low-noise amplifier (Mini-Circuits ZRL-1150-LN+) with an 80-dBm gain, and filtered through a 1-GHz band-pass filter (± 10 MHz) (KL Electronics 3C40-1000/T10-O/O) to suppress noise from unwanted frequency bands. A tunable attenuator (HP 8495B) was then used for linear calibration, and the processed signal was sent to a low-noise power detector (Mini-Circuits ZX47-60). The detector output voltage was measured by a lock-in amplifier (Stanford Research SR830) using the 1-kHz square wave as a reference. The lock-in readings were subsequently calibrated to a linear scale.

**Supplementary Information** includes:

- Training Loss Function
- Parameters and Digital Implementation for Numerical Analyses

**Figures**

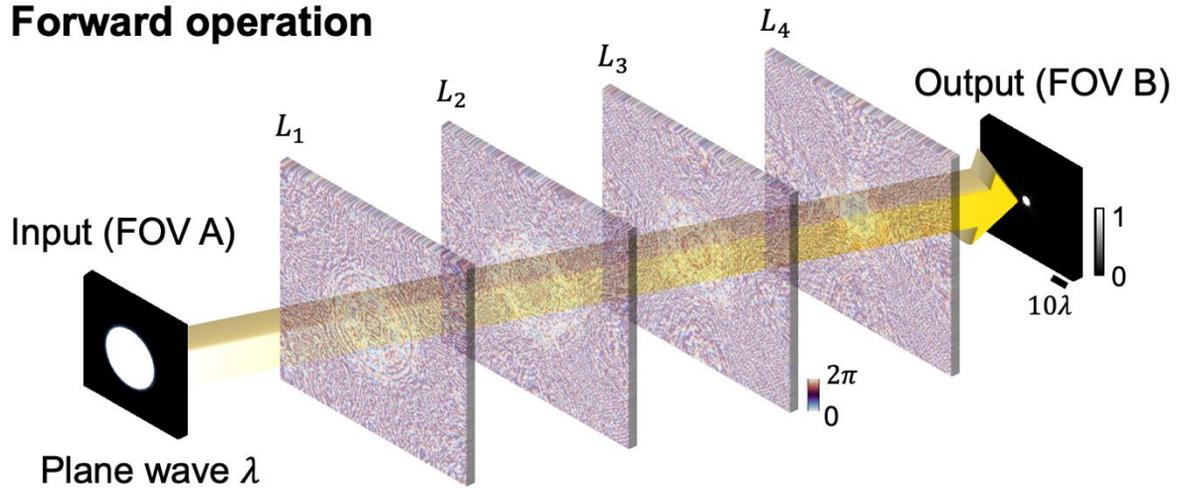

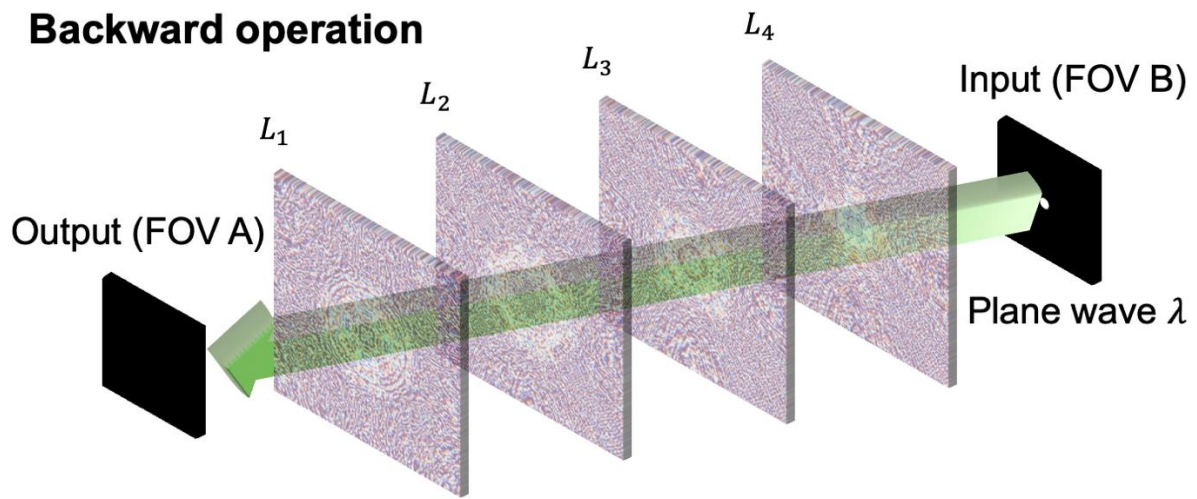

**Figure 1. Schematic of unidirectional focusing using a diffractive processor.** (**a**) The diffractive system focuses light efficiently in the forward direction (A→B). (**b**) The system blocks and scatters light in the backward direction (B→A).

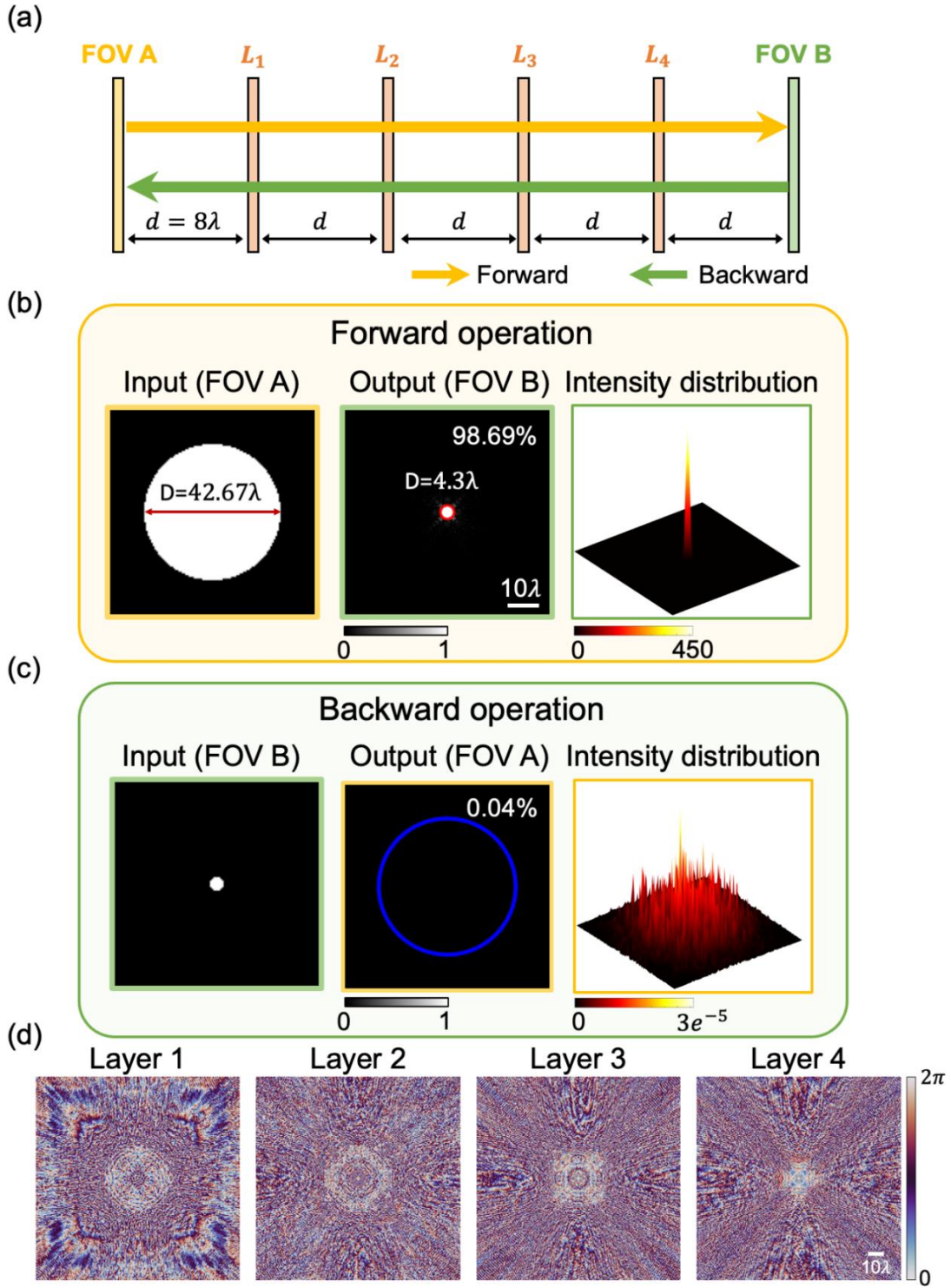

**Figure 2. Design schematics and numerical results of the diffractive unidirectional focusing system.** **(a)** Schematic layout of a four-layer unidirectional focusing system. **(b)** Forward-direction light focusing performance (A→B). **(c)** Backward-direction light blocking performance (B→A). **(d)** Phase profiles of the trained diffractive layers.

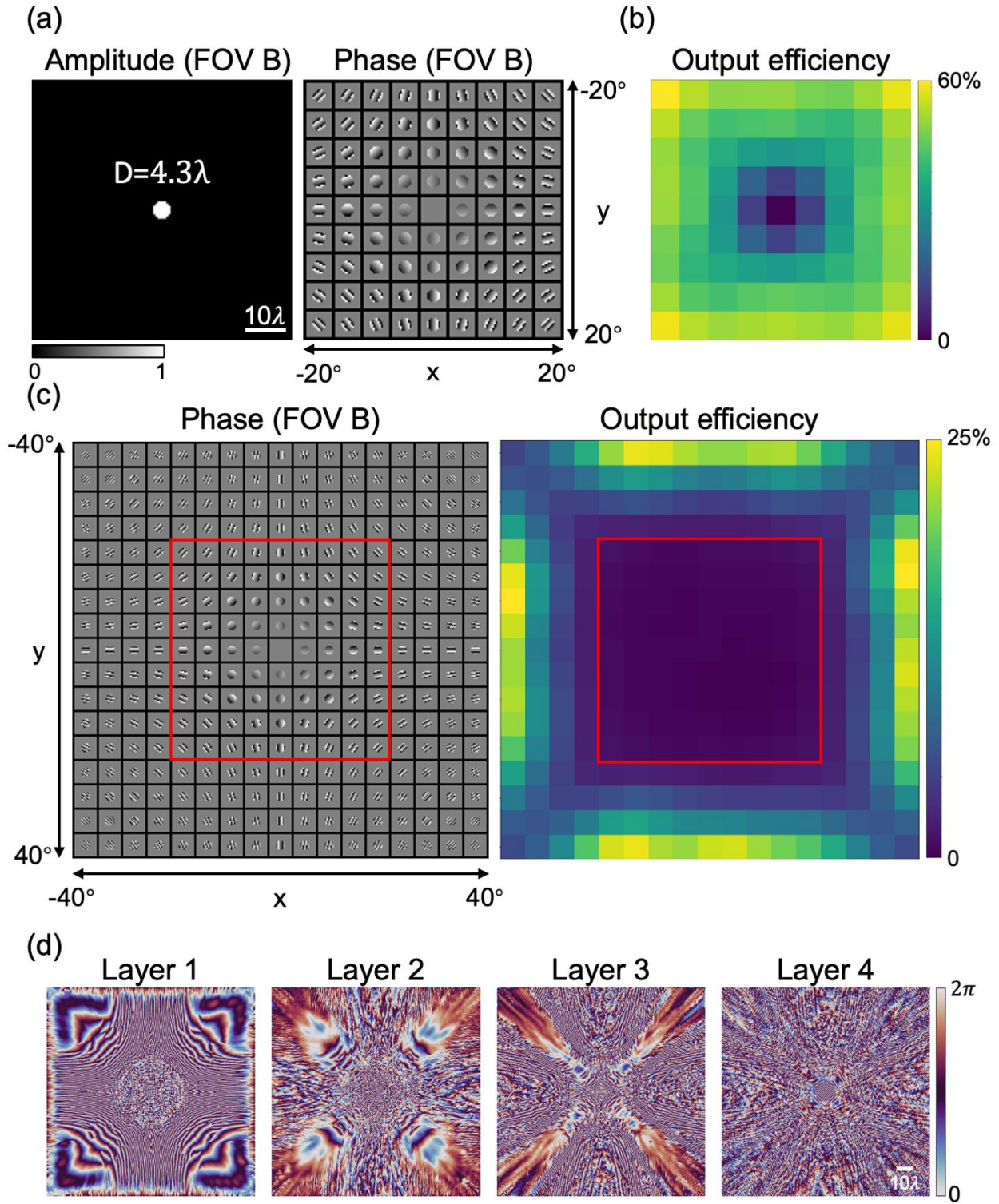

**Figure 3. Light-blocking performance under oblique illumination.** (a) Input amplitude and phase patterns forming the illumination patterns at various incidence angles at FOV B. (b) Output energy efficiency values of the unidirectional focusing system trained with normal incidence angle only. (c) Phase

profiles and the corresponding output energy efficiency values in the backward direction for the system tested under different incidence angles; the red box indicates the range of the incidence angles included during the training stage. **(d)** Phase profiles of the diffractive layers trained with multi-angle illumination.

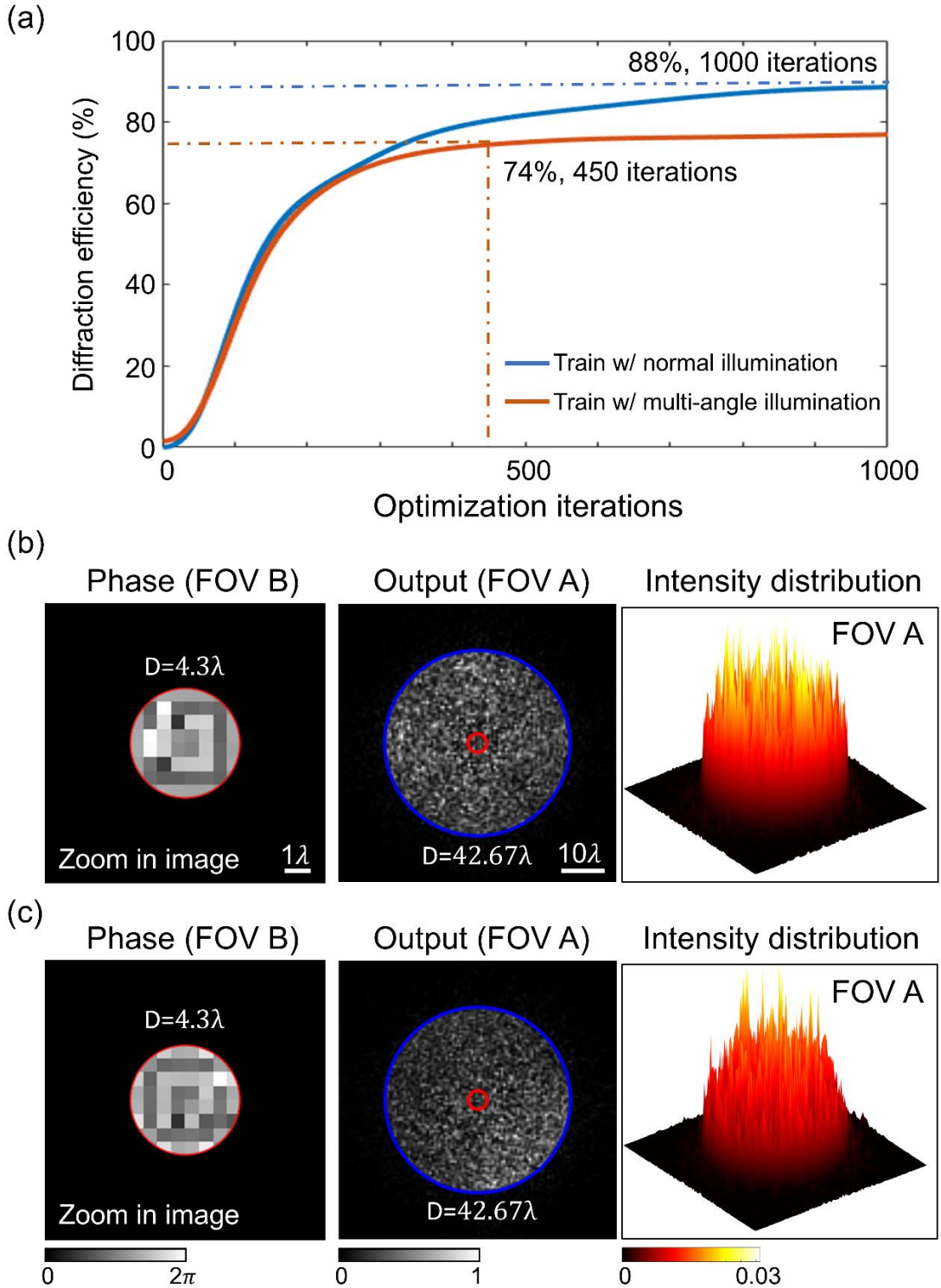

**Figure 4. Adversarial attacks on the unidirectional focusing system. (a)** An iterative adversarial attack on the unidirectional diffractive system is performed from outside using a "spy" detector from inside.

Optimization iterations of phase patterns in FOV B vs. the leakage energy efficiency in the backward direction. **(b)** Converged phase profiles in FOV B, output intensity in the backward direction (B→A), and the resulting intensity distribution at FOV A for the system trained only with normal incidence angle in the backward operation. **(c)** Corresponding results for the diffractive unidirectional system trained with multiple incidence angles in the backward operation. These adversarial attacks utilized direct access to the beam power at FOV A for iteratively optimizing the attacking wavefront profile from outside. Without this internal feedback using a spying detector at FOV A, such an adversarial attack cannot be successful from outside.

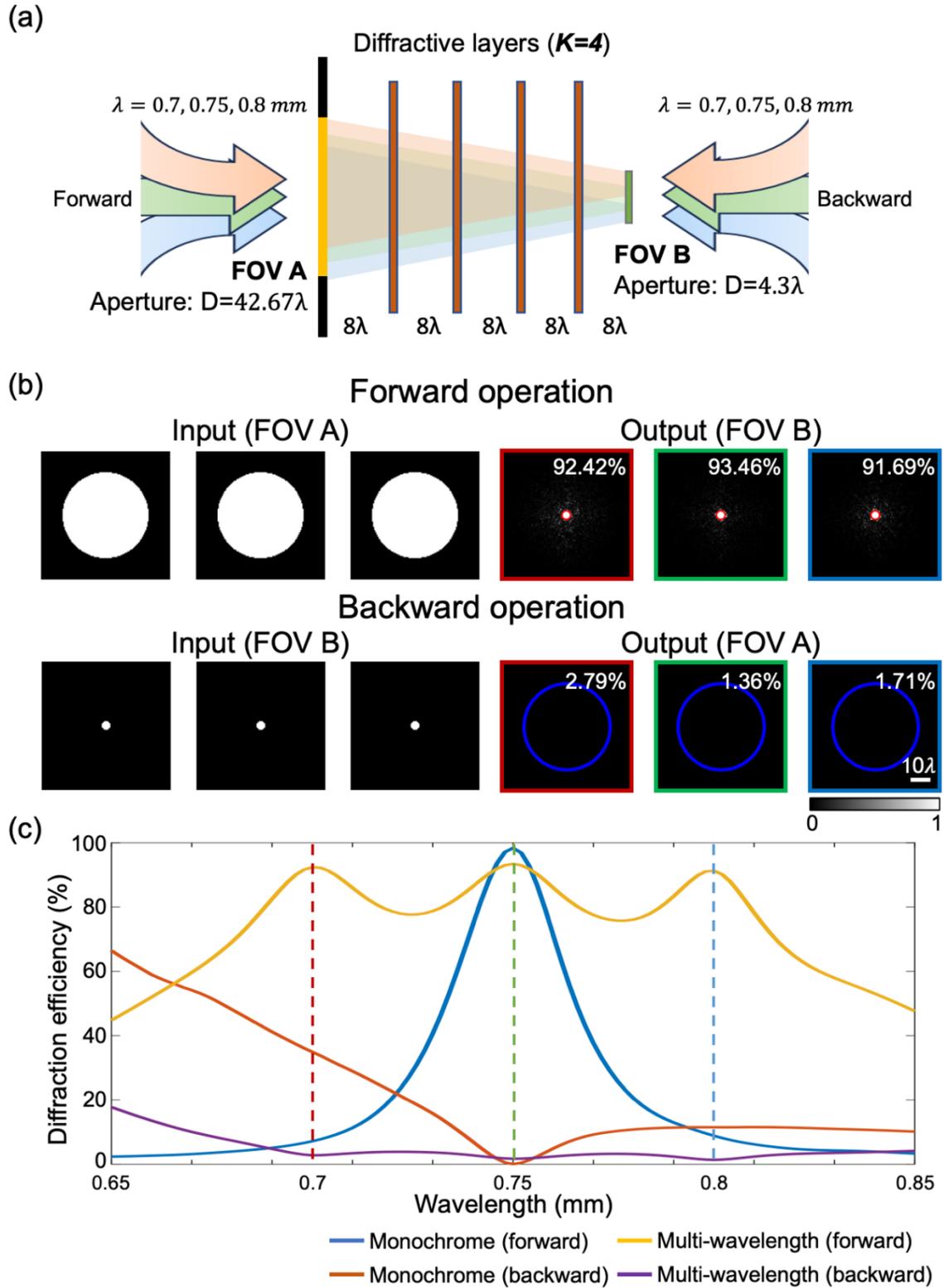

**Figure 5. Multi-wavelength unidirectional focusing systems and their spectral responses. (a)** Schematic representation of the multi-wavelength unidirectional focusing system. **(b)** Input and output

intensity profiles for the forward (A→B) and backward (B→A) directions. **(c)** Spectral responses of the diffractive unidirectional focusing systems trained with normal incidence compared to their counterparts trained with multi-angle incidence.

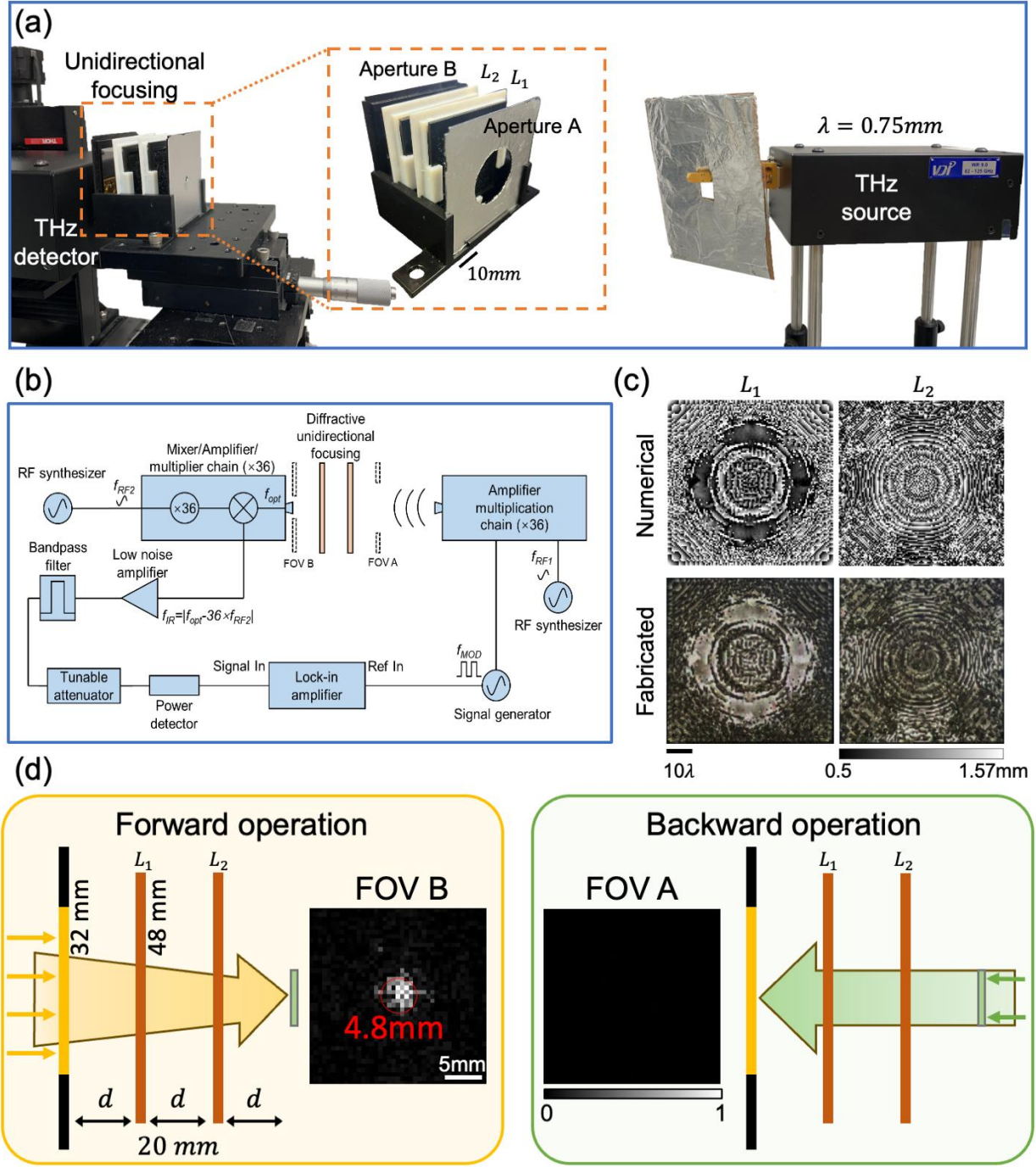

**Figure 6. Experimental setup and results for the diffractive unidirectional focusing system. (a)** Photograph of the experimental setup, including the fabricated diffractive system. **(b)** Schematic diagram of the continuous-wave terahertz setup. **(c)** Phase patterns of the learned diffractive layers alongside the photographs of the 3D-printed diffractive layers. **(d)** Experimental results for the forward focusing and backward beam blocking.